\documentclass[sigconf]{acmart}

\AtBeginDocument{%
  \providecommand\BibTeX{{%
    \normalfont B\kern-0.5em{\scshape i\kern-0.25em b}\kern-0.8em\TeX}}}

\setcopyright{acmcopyright}
\copyrightyear{2021} 
\acmYear{2021} 
\setcopyright{acmcopyright}\acmConference[Chinese CHI 2021]{The Ninth International Symposium of Chinese CHI}{October 16--17, 2021}{Online, Hong Kong}
\acmBooktitle{The Ninth International Symposium of Chinese CHI (Chinese CHI 2021), October 16--17, 2021, Online, Hong Kong}
\acmPrice{15.00}
\acmDOI{10.1145/3490355.3490357}
\acmISBN{978-1-4503-8695-1/21/10}

\usepackage{wrapfig}
\usepackage{balance}

\begin{document}

\title{Understanding Players' Interaction Patterns with Mobile Game App UI via Visualizations}


\author{Quan Li}
\affiliation{%
  \institution{School of Information Science and Technology\\ShanghaiTech University}
 \city{Shanghai}
  \country{China}}
\email{liquan@shanghaitech.edu.cn}

\author{Haipeng Zeng}
\affiliation{%
  \institution{School of Intelligent Systems Engineering\\Sun Yat-sen University}
  \city{Shenzhen}
  \country{China}}
  \email{zenghp5@mail.sysu.edu.cn}

\author{Zhenhui Peng}
\affiliation{%
 \institution{School of Artificial Intelligence\\Sun Yat-sen University}
 \city{Zhuhai}
 \country{China}}
 \email{zpengab@connect.ust.hk}

\author{Xiaojuan Ma}
\affiliation{%
  \institution{Department of Computer Science and Engineering\\The Hong Kong University of Science and Technology}
  \city{Hong Kong}
  \country{China}}
   \email{mxj@cse.ust.hk}


\begin{abstract}
Understanding how players interact with the mobile game app on smartphone devices is important for game experts to develop and refine their app products. Conventionally, the game experts achieve their purposes through intensive user studies with target players or iterative UI design processes, which can not capture interaction patterns of large-scale individual players. Visualizing the recorded logs of users' UI operations is a promising way for quantitatively understanding the interaction patterns. However, few visualization tools have been developed for mobile game app interaction, which is challenging with multi-touch dynamic operations and complex UI. In this work, we fill the gap by presenting a visualization approach that aims to understand players' interaction patterns in a multi-touch gaming app with more complex interactions supported by joysticks and a series of skill buttons. Particularly, we identify players' dynamic gesture patterns, inspect the similarities and differences of gesture behaviors, and explore the potential gaps between the current mobile game app UI design and the real-world practice of players. Three case studies indicate that our approach is promising and can be potentially complementary to theoretical UI designs for further research.
\end{abstract}




\maketitle

\section{Introduction}

\par Understanding players' interaction patterns in smartphone devices with mobile game apps is important for game experts to develop and refine their app products~\cite{punchoojit2017usability,folmer2007designing,bunian2017modeling}. For example, inspecting players' gesture sequences in a gaming app can help its game user experience (UX) researcher determine the players' gaming expertise and understand the players' engagement; understanding how players interact with mobile multi-touch devices can help the user interface (UI) designers verify if the intended UI of the app is convenient enough for the gamers, thus enabling appropriate modification and optimization of the original UI design.

\par Conventionally, game experts study players' interaction patterns with mobile game apps through intensive user studies with target players or iterative UI design processes to achieve their purposes~\cite{korhonen2006playability,park2013developing,hellsten2019iterative}. For example, UX experts would pay attention to the players' performance by extracting and analyzing their in-game behaviors to understand whether they are fully engaged in the game; UI design experts mainly propose mobile game UI designs based on some theoretical experiments and comparison with the designs of competitor products, trying to provide more user-friendly interface designs and ensure a smooth players' interaction. While these evaluation methods are informative and can potentially help identify the general patterns of the players' gaming interactions via mobile app UI, they could not provide quantitative understanding of user interactions with the app product itself.

\par One alternative way to assist game experts to study players' interactions with mobile game apps is to directly record the user interaction data from mobile devices by developing logging and analysis tools. Visualization is a promising way to enable game experts to dig into deeper the interaction data~\cite{harty2021logging,froehlich2007myexperience,krieter2018analyzing}. However, most existing visualization tools have long been developed for desktop-based interaction data and cannot be simply borrowed and applied in mobile game scenarios~\cite{li2016visual,li2019visualizing}. It becomes more challenging when the mobile game app supports intuitive direct-manipulation and multi-touch operations, e.g., the joysticks moving towards multiple directions, i.e., \textit{up}, \textit{down}, \textit{left}, \textit{right}, and different skill buttons possessed by the game character in the app, mimicking real-world metaphors. There have been some works seeking to measure and analyze low-level interaction metrics on mobile touch screen devices by inviting participants to perform various controlled tasks~\cite{anthony2012interaction,anthony2015children}. Nevertheless, most of them focus on aggregating primitive metrics such as task completion time and accuracy/error rate, which do not always capture high-level interaction issues that players may encounter in real-world usage scenarios. Some studies present visualization methods for low-level interaction logs to identify noteworthy patterns and potential breakdowns in interaction behaviors, e.g., elderly users' interaction behaviors~\cite{harada2013characteristics}, however, they target at a few specific application interfaces such as \textit{Phone}, \textit{Address Book} and \textit{Map} that do not require dynamic user operation. For mobile game that supports multi-touch options (e.g., with joysticks), players usually have more dynamic and complex interactions with the app UI, making it more challenging to visualize the interaction and identify the patterns behind.

\par To fill this gap, in this work, we present a visualization approach that aims to help game experts understand players' interaction patterns in the context of a multi-touch gaming app with joysticks and skill buttons and further improve its UI design. Particularly, we first identify players' dynamic gesture patterns that correspond to a series of interaction data. Then, we develop a visualization tool to help inspect the similarities and differences of gesture behaviors of players with different gaming expertise and explore the potential gaps between the current mobile game app UI design and real-world practice of players. We evaluate our approach on three cases (i.e., \textit{interaction skill comparison}, \textit{individual interaction skill}, and \textit{user interface design verification}). Both qualitative feedback and quantitative results of case studies suggest that our approach is promising and can be complementary to mobile game players' behavior understanding and theoretical mobile game UI designs.

\par Our contributions can be summarized as follows: 1) a gesture-based logging mechanism that comprehensively records user interaction and identifies players' gesture patterns; 2) a novel visualization approach that identifies the similarities and differences of high-level gesture behaviors on a touchable mobile device, and 3) case studies that provide both quantitative and empirical evidence to verify the efficacy of our approach and elicit promising UI design implications. In the following sections, we briefly survey the background and related work, followed by an observational study to identify the mobile interaction data characteristics and design requirements for the proposed visualization approach. Then, we carry out three case studies to verify our approach. Finally, we conclude our work with discussions and limitations and shadow the potential design implications discovered through our approach.

\section{Related Work}
\par Literature that overlaps with this work can be categorized into three groups: mobile interaction data analysis, assessment of mobile UI design, and gesture data analysis.

\subsection{Mobile Interaction Data Analysis}
\par User behavior analysis has been intensively studied in the game domain~\cite{li2016visual,kang2013online,lee2014learning}. Li et al. analyzed players' actions and game events to understand reasons behind snowballing and comeback in MOBA games~\cite{li2016visual}. These studies mainly focus on in-game players' behavior analysis instead of their interaction with the game devices. The most similar studies come from web search~\cite{blascheck2014state,kim2015eye,huang2012user}, where statistical analyses of e.g., eye-tracking or mouse cursor data mainly provide quantitative results. Visualization techniques are developed to allow researchers to analyze different levels and aspects of data in an explorative manner, which can be categorized into three main classes, namely, point-based, area-of-interest-based, and approaches that combine both techniques~\cite{blascheck2014state}. Among these methods, the aggregation of fixations over time or participants is known as a heat map that summarizes illustrations of the analyzed data and can be found in numerous publications. However, many methods proposed for desktop website analysis cannot be simply applied to mobile game apps since these methods are based on single-point interaction such as mouse or eye movements while mobile devices support intuitive direct-manipulation and multi-touch operations~\cite{humayoun2017heuristics}.
\par Timelines are frequently used to represent the interaction information. Previous solutions have provided static and limited representations of them~\cite{carta2011support}. A new solution was designed to represent and manipulate timelines, with events represented by a level and a small colored circle~\cite{burzacca2013analysis}. It also includes vertical black lines among events that indicate a page change, which provides effective interactive dynamic representations of user sessions. Nebeling et al. presented \textit{W3Touch} to collect user performance data for different mobile device characteristics and identify potential design problems in touch interaction~\cite{nebeling2013w3touch}. Guo et al. conducted an in-depth study on modeling interactions in a touch-enabled device to improve web search ranking~\cite{guo2013mining}. They evaluated various touch interactions on a smartphone as implicit relevance feedback and compared these interactions with the corresponding fine-grained interactions in a desktop computer with a mouse and keyboard as primary input devices. Bragdon et al. investigated the effect of situational impairments on touch-screen interaction and probed several design factors of touch-screen gestures under various levels of environmental demand on attention compared with the status-quo approach of soft buttons~\cite{bragdon2011experimental}. To date, however, few empirical studies have been conducted on mining touch interactions with mobile game apps to understand players' behaviors and further suggest mobile application UI designs via a visual analytics approach.

\subsection{Assessment of Mobile UI Design}
\par Many mobile applications, particularly game apps such as ARPG (Action Role Playing Game), introduce gesture operation to control the game character in the game app freely~\cite{hesenius2014automating,ruiz2011user}. However, no single gesture operation can resolve all the interaction issues in the mobile game application scenarios due to different screen sizes, individual behaviors, and the form of different controls used. Besides, gesture operation requires extensive learning. Thus, most existing mobile game applications, particularly role-playing games, adopt a virtual joystick and skill buttons for players' interaction~\cite{baldauf2015investigating}.

\par Previous studies that focus on the assessment of mobile UI design can be summarized into two categories, qualitative methods and quantitative methods. For example, when designing mobile app interfaces, targets are generally large to make it easy for users to tap~\cite{anthony2012finger}. The iPhone Human Interface Guidelines of Apple recommend a minimum target size of $44$ pixels wide * $44$ pixels tall~\cite{clark2010tapworthy}. The Windows Phone UI Design and Interaction Guide of Microsoft suggests a touch target size of $34$ pixels with a minimum touch target size of $26$ pixels * $26$ pixels. The developer guidelines of Nokia suggest that the target size should not be smaller than $1$ cm * $1$ cm or $28$ pixels * $28$ pixels. Although these guidelines provide a general measurement for touch targets, they are inconsistent and do not consider the actual varying sizes of human fingers. In fact, the suggested sizes are significantly smaller than the average finger, which may lead to many interaction issues for mobile app users. Hrvoje et al. presented a set of five techniques, called dual finger selections, which leveraged recent developments in multitouch sensitive displays to help users select extremely small targets~\cite{benko2006precise}. The UED team of Taobao researched to determine hotspots and dead-ends and to identify the control size using the thumb\footnote{http://www.woshipm.com/pd/1609.html}; they concluded that the minimum target size should be $11$ mm by single thumb operation to achieve an average accuracy higher than 95\%. However, most existing works have focused on the vertical screen mode by single-hand operation, and only a few have discussed the landscape mode using both hands. In this work, we focus on the assessment of virtual joystick and skill buttons by leveraging players' interaction data with the mobile game app to analyze the triggering and moving ranges of virtual joystick and skill set areas.

\par Regarding the quantitative methods, researchers have been working on modeling user perception and subjective feedback of user interface, e.g., the judgment of aesthetics~\cite{miniukovich2015computation}, visual diversity~\cite{banovic2019computational}, brand perception~\cite{wu2019understanding}, and user engagement~\cite{wu2020predicting}. Typically, a set of visual descriptors would be compiled to depict a UI page in terms of e.g., color, texture, and organization. Specifically, user perception data are collected at scale and corresponding models are constructed based on some hand-crafted features~\cite{wu2019understanding}. However, feature engineering cannot ensure a comprehensive description of all the aspects of UI. Recently, deep learning has demonstrated its decent performance on learning representative features~\cite{krizhevsky2012imagenet}. For example, a convolutional neural network is adopted to predict the look and feel of certain graphic designs. Wu et al. leveraged deep learning models to predict user engagement level on animation of user interfaces~\cite{wu2020predicting}. Similarly, perceived tappability of interfaces~\cite{2019Modeling} and user performance of menu selection~\cite{10.1145/3173574.3173603} can also be predicted with the assistance of deep learning methods. However, the existing studies provide prediction scores of user perception towards different UI designs while in our work, we study how players experience in real-world practice via the current design of mobile game UI by visually analyzing their interaction patterns and shadow the design implication of mobile game UI interface.

\subsection{Gesture Data Analysis}
\par Owing to the pervasiveness of multi-touch devices and wide usage of pen manipulation or finger gestures, a great number of researchers have conducted on stroke gestures analysis generated by users~\cite{zhai2012foundational,10.1145/2797138,10.1145/3290605.3300445,10.1145/3229434.3229478}. Most of the related existing works focus on gesture recognition, which matches gestures with target gestures in the template library based on their similarity. For example, Wobbrock et al. developed an economical gesture recognizer called \$1, which is easy to incorporate gestures into UI prototypes~\cite{10.1145/1294211.1294238}. They employed a novel gesture recognition procedure, including re-sampling, rotating, scaling and matching without using libraries, toolkits or training. Yang Li developed a single-stroke and template-based gesture recognizer, which calculates a minimum angular distance to measure similarity between gestures~\cite{10.1145/1753326.1753654}. Ouyang et al. presented a gesture short-cut method called \textit{Gesture Marks}, which enables users to use gestures to access websites and applications without having to define them first~\cite{10.1145/2207676.2208695}. In order to understand the articulation patterns of user gestures, Anthony et al. studied the consistency of gestures both between-users and within-users and some interesting patterns have been revealed~\cite{2013Anthony}. Some works conduct research on different users, such as children~\cite{anthony2012interaction,anthony2015children}, elderly people~\cite{2018LinYuHao}, and people with motor impairments~\cite{10.1145/3290605.3300445}, aiming to improve user experiences on mobile device interactions.
\par In this work, instead of gesture recognition, we focus on analyzing user behaviors to find similar and preferred stroke gestures, i.e. similar operations generated by players when playing games. Considering different features of stroke gestures in terms of position, direction, shape and so on, we cluster stroke gestures to reveal users' common behaviors by resampling a stroke gesture as a point path vector, followed by a definition of a distance function and clustering algorithms.

\section{Observational Study}
\par To understand the mobile game app interaction data characteristics and identify the design requirements of our visualization approach, we worked with a team of experts from an Internet game company, including two UX analysts (E.1, male, age: $24$ and E.2, female, age: $26$), one data analyst (E.3, male, age: $25$) and two game UI designers (E.4, female, age: $24$ and E.5, female, age: $25$), a typical setup for a game UX team in the company. All of them have been in the game industry for more than two years. To obtain an understanding of players' interaction patterns and experiences with the mobile game app, the game experts have different responsibilities. Particularly, E.1 and E.2 would conduct two main approaches, namely, in-game interaction data analysis with the help of E.3 and subjective interview with the testing players to understand their interaction patterns and provide UI design suggestions for E.4 and E.5. To obtain detailed information of the current practice of the game experts, with consent, we shadowed the team's daily working process, including videotaping how they observed players experiencing the game, conducted testing experiments, and on-site interviews with the players. Later, we carried out retrospective analysis together with the game team on their conventional practices.

\par \textbf{Participants.} The game experts recruited $18$ participants ($5$ female, avg. age: $24$) from a local university. They were undergraduate and postgraduate students. Some of them are novice players, ensuring that the study of players' interaction applies to different expertise of target users.

\par \textbf{Procedure.} In the prior study, the participants were first asked to complete a task by a mobile app similar to the game of Whack-A-Mole\footnote{https://wordwall.net/about/template/whack-a-mole} using a mobile phone device with the size of $11.07$ cm * $6.23$ cm and the resolution of $1920$ * $1080$p. Particularly, the mobile screen is divided into multiple small squares, of which a random square turns red, requiring the participants to accurately click on within $1$ second. In this experiment, the size of the square and its position were randomly combined, and the participants needed to click $100$ times consecutively and this lasts for about one and a half minutes. The participants held their hands horizontally and clicked on the screen with their thumbs. The square size was designed as a variable with the side length taking from $6$mm to $15$mm. The keystroke time is defined as the time from the appearance of the red square to the time when the player clicks on the screen and the game experts calculated the distance from the red square to the lower-left corner of the screen. The game experts distinguished the square size and position for data statistics.

\begin{figure}[h]
\includegraphics[width=\linewidth]{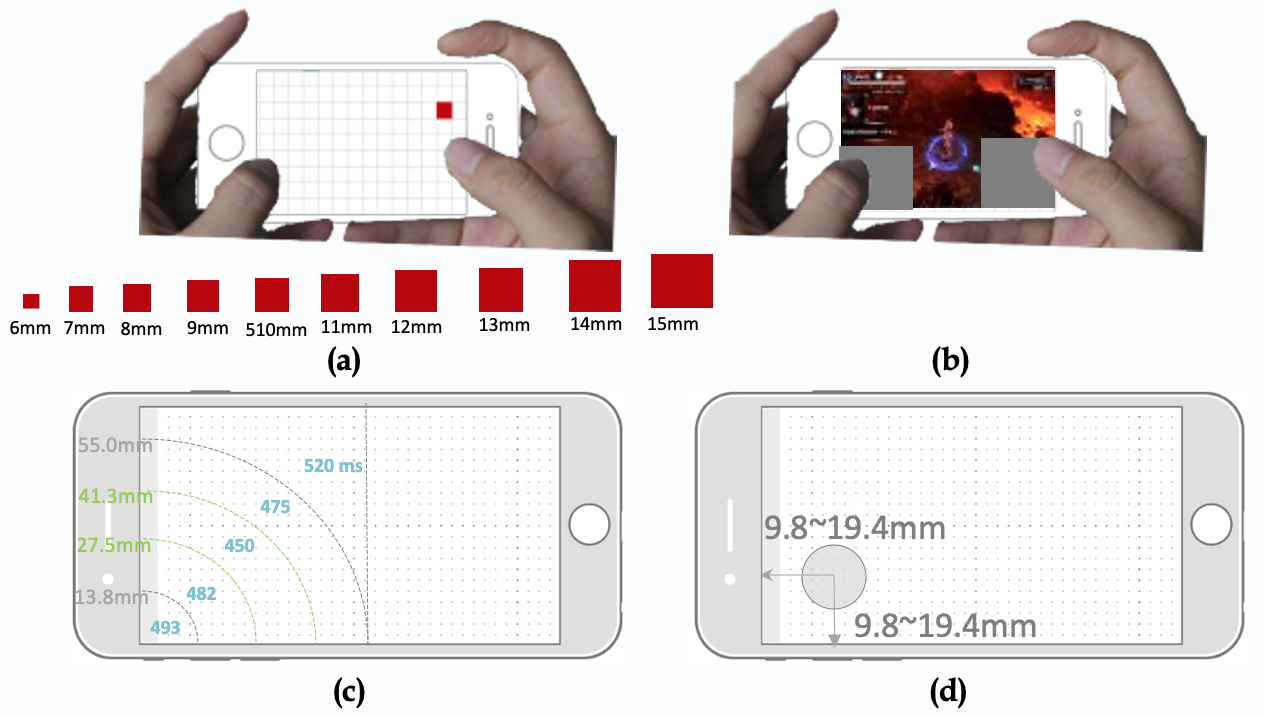}
\vspace{-6mm}
\caption{(a) The participants were required to hold the phone horizontally and click on the randomly appeared red square with different sizes with their thumbs on the screen; (b) The game experts covered the main operation regions of the mobile game app interface to ensure that the participants operate based on their own habits for an objective evaluation; (c) Identifying the comfortable clicking zones for the left thumb, i.e., the shortest response time; (d) Translating the coordinates to the distances between the center of the joystick and the left and bottom of the screen.}
\label{fig:procedure1}
\vspace{-3mm}
\end{figure}

\par \textbf{Result Analysis.} To match players' real-world usage scenarios and regarding the size and pixel adaption of different mobile phones, the game experts take ``mm'' as the measurement unit instead of pixel. As shown in \autoref{fig:procedure1}(c), the most comfortable zone for the left thumb's clicking (i.e., the shortest reaction time) is the area with a radius of $27.5$ -- $41.3$ mm. Considering the occlusion of the combat interface when the joystick is moved to the middle of the screen, the UI designers maintained that the secondary comfort zone ($13.8$ -- $27.5$ mm) is recommended as the design area. The designers further converted to the distances between the center of the joystick and the left and bottom of the screen, i.e., $9.8$ mm -- $19.4$ mm. The experts also investigated other game competitors' UI designs and identified that most of the center of the joystick keeps at least a $10$ mm distance from the left and lower corners of the screen (\autoref{fig:procedure1}(d)). Note that the results are based on the randomly appearing square button and the focus area is the recommended design area, i.e., the lower left part of the screen when holding the phone horizontally. Specifically, when the side length is set to $12$ mm, the accuracy of square clicking\footnote{click accuracy: in a single response, participants correctly hit the target as the end. If the number of hit is greater than 1, the response is considered to be a failure. That is, the click accuracy is the percentage of the number of times that the finger successfully hit the stimulating red square for the first time to the total number of the red squares} can be larger than $90$\% (Table~\ref{table1}). The joystick size of game competitors is basically consistent with experimental results, i.e., between $11.9$ and $14.6$ mm. 

\begin{figure}[h]
\includegraphics[width=\linewidth]{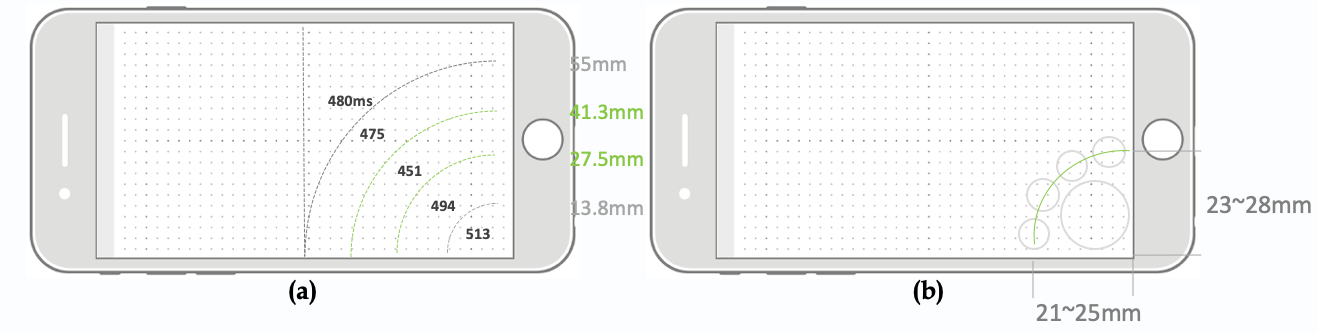}
\vspace{-6mm}
\caption{Testing results of the comfortable zone for the right thumb's clicking.}
\label{fig:procedure2}
\vspace{-3mm}
\end{figure}

\par Similarly, the UX experts found that the most comfortable zone for the right thumb's clicking (i.e., the shortest reaction time) is also the area with a radius of $27.5$ -- $41.3$ mm, as shown in \autoref{fig:procedure2}(a). And the secondary comfort zone, i.e., $13.8$ -- $27.5$ mm is also recommended as the design area. Although $41.3$ -- $55$ mm is also a comfortable zone for players to operate, it can easily obscure the screen. UI design experts suggested that as the $41.3$ mm arc is close to the edge of the screen, some heavy skills can be placed here, e.g., ultra-low frequency skill buttons. The game experts also surveyed other game competitors' low-frequency skill buttons and concluded that their positions are close to the $27.5$ mm arc, i.e., the most comfortable area (\autoref{fig:procedure2}(b)).

\par Following a similar procedure, the game experts found that when the thumb's clicking range is within the comfort area recommended above, 90\% accuracy can be ensured if the diameter of the Normal Attack button is over $9$ mm and the diameter of the other skill buttons is above $8$ mm (\autoref{table1}). The survey of game competitors also confirms that the diameter of skill buttons is within $7$ to $11$ mm and the Normal Attack button is within $11$ to $15$ mm.

\begin{table}[h]
\centering
\caption{Click accuracy of participants.}
\label{table1}
\resizebox{\linewidth}{!}{%
\begin{tabular}{|c|c|c|c|c|c|c|}
\hline
{Size (mm)}                                                              & 7      & 8      & 9      & 10     & 11     & 12      \\ \hline
\begin{tabular}[c]{@{}c@{}}Normal Attack\\ (13.75 -- 27.5 mm)\end{tabular} & 77.3\% & 77.8\% & 94.4\% & 85.7\% & 87.5\% & 100.0\% \\ \hline
\begin{tabular}[c]{@{}c@{}}Skills\\ (27.5 -- 41.25 mm)\end{tabular}        & 63.0\% & 90.6\% & 92.3\% & 90.5\% & 80.0\% & 100.0\% \\ \hline
\end{tabular}%
}
\end{table}

\par The results of the above experiments provide some initial guidelines for the UI design of their mobile game app. However, UI design experts (E.4 -- 5) commented that ``\textit{although in general, they are consistent with the survey results of other game competitors, they can be quite rough and general.}'' E.1 further commented that ``\textit{this experiment requires a high degree of concentration but in reality, players are playing the game in a more relaxed mood.}'' In other words, the interaction characteristics of a certain mobile game app are not fully considered during the testing when the participants were conducting their actions and they are in a state of tension. UI design experts, therefore, envisioned a more customized and natural way to learn the players' interaction patterns with the mobile devices for inferring the design suggestions for the joystick and skill buttons. On the other hand, UX experts mainly studied the players' in-game performance. However, in-game metrics reflects the level of players' performance but cannot explain the reasons behind their performances, ``\textit{while a bad interaction with the game app would certainly lead to poor performance of players, this observation cannot be easily captured by the in-game metrics,}'' said E.3. That is, a good way to identify similarities and differences among players' interactions with the mobile game app is still missing for the UX experts.

\par To ensure that the ontological structure of our approach fits well into domain tasks, we interviewed the game experts (E.1 -- 5) to identify the experts' primary concerns about the analysis of players' interaction patterns with playing the mobile game app and potential obstacles in their path to efficiently obtaining insights. At the end of the interviews, the need for a gesture-based visualization approach to ground the team's conversation with assessing mobile interaction patterns emerged as key themes among the feedback collected. Despite differences in individual expectations for such approaches, certain requirements were expressed across the board.
\par \textbf{R.1 Distinguishing gesture behaviors from a spatiotemporal perspective.} Conventionally, the game experts leveraged heat map visualizations to observe the distribution of clicking spots, regardless of interactions caused by different fingers' interactions, which cannot show the quantitative information. Furthermore, the collective heat map distribution cannot provide more details of different gesture behaviors that may occur at different timestamps and positions, failing to shed lights upon players' interaction patterns. Therefore, the game experts wished to distinguish different gesture behaviors from a spatiotemporal perspective.

\par \textbf{R.2 Inspecting behavior differences and similarities among players.} One concern of the game experts was that the interaction patterns cannot be easily inspected only by leveraging the in-game metrics. For example, the game experts wished to know ``\textit{what the common interactions with the mobile device is among the players and what the difference is}'', thus allowing them to understand how the operation skill may influence their in-game performance, which can be complementary to the performance of in-game behaviors.

\par \textbf{R.3 Identifying interaction areas with appropriate scales and positions.} UI design experts typically focus on three aspects of designs, i.e., style design, scale design, and position design. While style design is reflected in the interactive elements, caters to the gameplay experience, and has sufficient feedback to the player's operation behavior, players' interaction patterns can be largely influenced by the scale and position designs. Therefore, the game experts, especially the UI design experts, wanted to identify the appropriate interaction areas in terms of scales and positions that can reflect the real-world gameplay interaction experiences accurately.

\begin{figure*}[h]
\includegraphics[width=\linewidth]{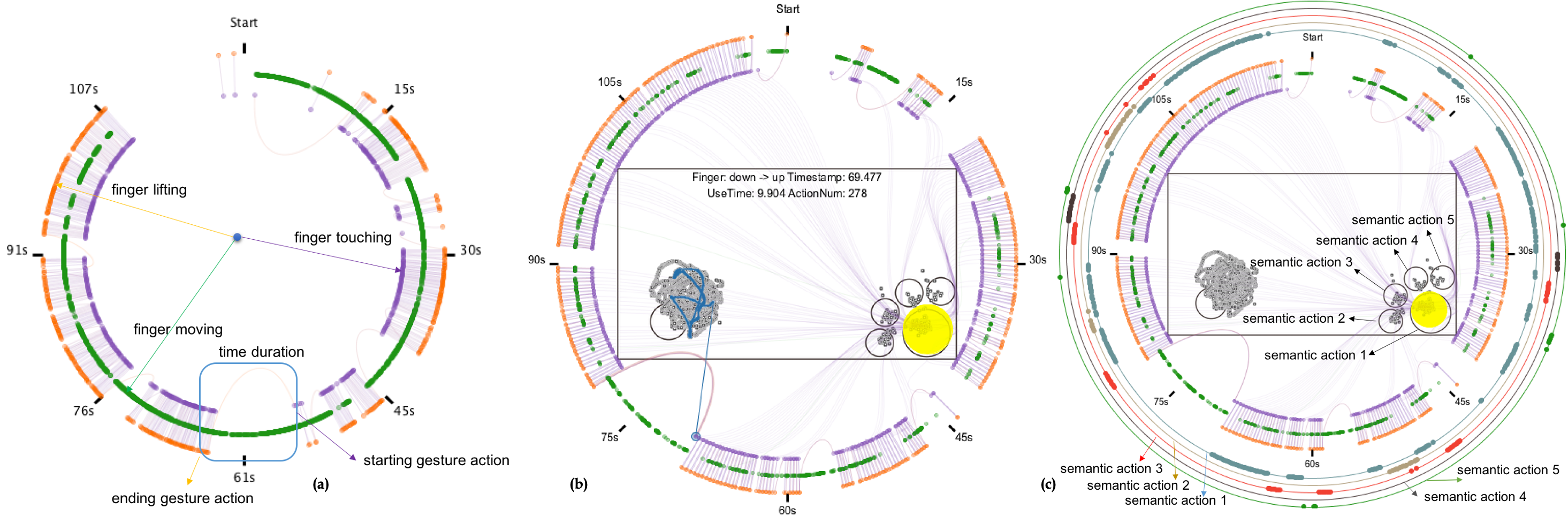}
\vspace{-6mm}
\caption{(a) Three different basic finger actions are placed on three concentric circles: finger touching on the most inner circle, finger moving on the second inner circle, and finger lifting on the most outer circle. All actions are distributed in the clockwise direction, as shown by the timestamps aside. A complete gesture starts with a finger touching (``starting'' point), followed by several finger movings, and ends with a finger lifting (``ending'' point). We link the ``starting'' point to the ``ending'' point by a curve, as indicated by the blue rectangle. The duration of the gesture can be obtained by calculating the timestamp difference between the ``starting'' point and the ``ending'' point; (b) Select an area (the yellow circle) and its corresponding finger actions distributed on the concentric circles will be linked by curves; hover on a finger touching action and the gesture trajectory will be displayed (the blue point and the blue trajectory); (c) Five semantic interactions are mapped onto five outer semantic circles. Users can self-define the semantic areas by drawing a circle to surround them and assign it certain semantic information. Different skills are then mapped onto the corresponding outer concentric circles.}
  \label{fig:design1}
  \vspace{-3mm}
\end{figure*}

\section{Approach}
\par In this section, we first illustrate how we collect the gesture-based interaction data and then introduce our visualizations to help game experts understand the mobile game app interaction patterns.

\subsection{Gesture-based Logging Mechanism}
\par We have developed an application-independent Android program that can interact with the mobile OS. By detecting every touch event and retrieving the screen coordinates with the timestamp of the touchable screen, we can log all the touchable-screen actions through multiple functions, e.g., ACTION\_DOWN (touch the screen), ACTION\_MOVE (move on the screen), and ACTION\_UP (leave the screen). We consider the players' interaction data as high-level gestures, which are the trajectories of players' fingers on the multi-touch screen that can be recorded as a series of points generated by the same finger of the player in one session of interaction. Particularly, when the player places a finger on the screen, a starting point $P_0$ is recorded. The player then moves this finger to other places, generating several corresponding points ($P_1$, $P_2$, ..., $P_n$). When this interaction session terminates, the player raises his/her finger and ends the recording of the gesture. We define the gesture $g_k$ = [$P_0$, $P_1$, ..., $P_i$, ..., $P_n$], where point $P_i$ is described in terms of ($x_i$, $y_i$) with corresponding timestamp $T_i$ indicating the corresponding time-lapse from the beginning. Each gesture trajectory has an associated id, which records the current session the program identifies during the interactions.

\subsection{Gesture-based Visualization}
\par The basic design principle behind our approach is leveraging or augmenting familiar visual metaphors to enable game experts to focus on analysis~\cite{8802454}. Considering the gesture data being analyzed and the above-mentioned requirements, we visually encode the data in a manner that ensures that the patterns and outliers are easily distinctive without overwhelming the analysts.

\par We define a gesture as a series of finger actions, typically starting with a ``finger down'', followed by several ``finger moving'', and ending up with ``finger lifting''. We adopt a timeline-like metaphor to align all the actions of the fingers in a radial layout, presenting the overview of the distribution of finger actions intuitively, as shown in \autoref{fig:design1}. We choose a radial layout because it allows users to analyze high-level interactions (e.g., periodic behaviors) and compare the gesture patterns in different stages in a more concentrated manner. Meanwhile, the mobile game interface could be placed inside the radial circles, helping analysts better link temporal events with the corresponding interaction dots and spatial trajectories on the screen. We apply three concentric circles to denote the three fundamental actions of fingers, namely, ``finger touching'' in the most inner circle, ``finger moving'' in the second inner circle, and ``finger lifting'' in the outermost circle (R.1). The spatial position of each finger action is determined by its timestamp and the entire recorded period is considered in a clockwise direction. One complete and continuous gesture is represented by a Bezier curve that links the starting action (finger touching) and the ending action (finger lifting) with a series of finger movements distributed between the two actions. We also encode gesture durations into the parameters of two controlling points of the Bezier curve, i.e., the height of the curve corresponds to the related gesture duration. We also provide heat map visualization as a qualitative complementary to the proposed gesture visualization.

\par In addition, to encode the gestures to concentric circles, we filter out gesture trajectories based on spatial locations, e.g., we can visualize the trajectories that correspond to a particular area of the mobile app interface (R.1). Particularly, we design a spatial-based query to link the trajectories to the corresponding finger actions along the concentric circles (see the selected area). We directly integrate the UI design into the visualization. On the left side of the view, the interaction within the area controls the orientation and motion of the game character in the app, and the right side, i.e., the five rounded areas correspond to different skills of the game character. \autoref{fig:design1}(b) presents the use of the spatial-based query for a relatively long duration of operation, which corresponds to a gesture trajectory. Most of the finger movements occur on the left side, whereas quick taps happen on the right side.

\par Since the skill button interaction represents certain semantic information, i.e., different skill button indicates different operations on the game character, we can define the high-level interactions in the skill semantic space. As shown in \autoref{fig:design1}(c), analysts can self-define a region, e.g., a circle to surround an area and assign certain semantic information to the area, and the corresponding finger actions within the area would be automatically mapped onto the semantic axes on the outermost semantic circles, which supports the interactive exploration from the skill semantic perspective (R.2).

\par
\begin{figure}{h}
\centering
\includegraphics[width=\linewidth]{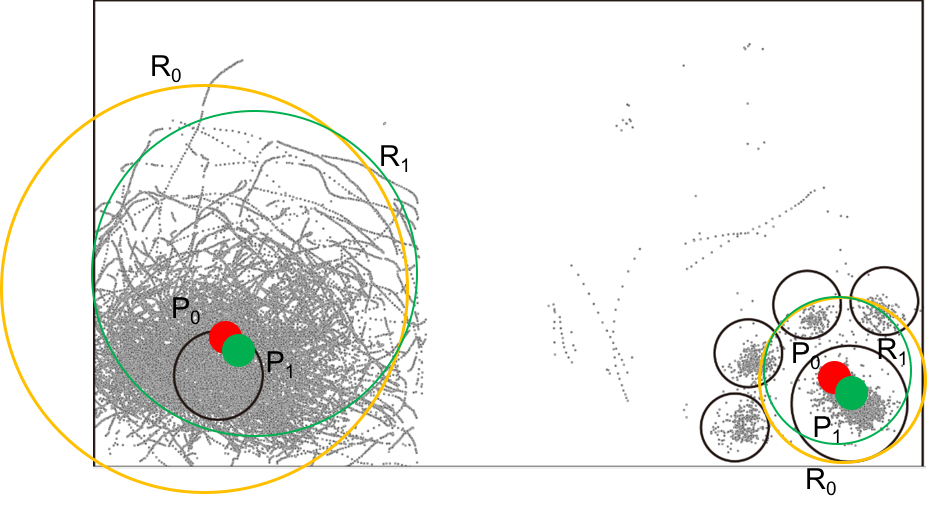}
\vspace{-6mm}
\caption{The selected area is represented by yellow and the original clustering center is indicated by a red dot; when choosing the parameter of confidence coefficient as 95\%, the new clustering center is shown by a green dot and the new adjusted area is represented by a green zone.}
\label{fig:cluster_design}
  \vspace{-3mm}
\end{figure}

 Conventionally, to identify the typical interaction patterns on the mobile game app screen, the heat map is intensively used. One obvious advantage is the lack of quantitative feedback by using a heat map that only conveys a sense of qualitative density information. After discussion with the game experts, we propose an interactive clustering method based on the interaction data (R.3). Particularly, we allow analysts to select the original interaction dots with a certain radius $R_0$ and the system automatically calculates the clustering center $P_0$ of the selected interaction dots and choose the area with different confidence coefficients, i.e., identify the most appropriate center and size of the area that meets certain confidence coefficients. In this way, a new clustering center $P_1$ could be generated. We further sort the interaction dots based on the distance between $P_1$ and the interaction dot $d_k$ and the longest distance is considered as the new radius $R_1$ (\autoref{fig:cluster_design}).

\par Once we determine the region for a certain confidence coefficient, we then extract the gestures within the region to understand the underlying semantics of the gestures and reveal the common interaction behaviors of players. Note that different gestures contain various number of interaction dots, we need to resample gestures to ensure that all the gesture trajectories are directly comparable. Given a gesture $g_k$, we assume its original number of interaction dots is $M$ and the objective is to sample $N$ interaction dots from $g_k$ with evenly spaced sampling. Specifically, the original gesture is represented as $g_k = [P_0, ..., P_i, ..., P_{M-1}]$ and $P_i = (x_i, y_i)$ and after sampling, we get a new gesture $v_k = [P_0^s, P_1^s, ..., P_i^s, ..., P_{N-1}^s]$, where $P_0 = P_0^s$ and $P_{M-1} = P_{N-1}^s$. They are subjective to 
\begin{equation}
dist(P_i^s, P_{i+1}^s) = \frac{\sum_{i=0}^{M-2}\sqrt{(x_{i+1} - x_i)^2 + (y_{i+1} - y_i)^2}}{(N-1)}.
\label{eq:1}
\end{equation}
In other words, we add the first interaction dot $P_0$, and then add a new dot sequentially until $v_k$ covers $N - 1$ dots, followed by adding the last interaction dot. Those new interaction dots is generated from the original gesture $g_k$ by using linear interpolation. The next step is to measure the distance between two gestures. Particularly, given two gesture sampling vectors: $v_k(P_i = (x_{ki}, y_{ki}))$, $v_h(P_i = (x_{hi}, y_{hi}))$, and considering the absolute distance and directions, we combine two distance functions, i.e., Euclid Distance and Cosine Distance with adjustable weights, and use K-Means to all sampling gesture vectors. The distance functions are defined as Euclid Distance:
\begin{equation}
d_1 = \sqrt{\sum_0^{N-1}(x_{hi} - x_{ki})^2 + (y_{hi} - y_{ki})^2}
\label{eq:2}
\end{equation}
and Cosine Distance
\begin{equation}
d_2 = v_k \cdot v_h/|v_k||v_h|,
\label{eq:3}
\end{equation}
where 
\begin{equation}
v_k \cdot v_h = \sum_0^{N-1}(x_{ki}x_{hi} + y_{ki}y_{hi})
\label{eq:4}
\end{equation}
and
\begin{equation}
|v_k||v_h| = \sqrt{\sum_{i=0}^{N-1}(x_{ki}^2 + y_{ki}^2)}\sqrt{\sum_{i=0}^{N-1}(x_{hi}^2 + y_{hi}^2)}.
\label{eq:5}
\end{equation}

\section{Case Study}
\par To evaluate our approach, we conduct several case studies, in which the previous $18$ participants are asked to play a mobile game to collect their interaction data for our analysis. Then, we present our visualization approach to the game experts to evaluate the efficacy of our design. Particularly, the evaluation is conducted in the following three cases to examine whether our visualization approach fulfills the aforementioned requirements.
\subsection{Participants and Procedures}
\par The education backgrounds of the above-mentioned participants range from computer science, electronic engineering to art designs. We collect information about the participants' mobile phone usage, including their mobile phone operating systems, the size of their mobile screens, and the frequency of playing mobile games per week. All of them have the experience of using both \textit{IOS} and \textit{Android} mobile operating systems. The size of the mobile phone screens they use ranges from $3.1$ to $3.5$ inch to above $5.0$ inch. Regarding mobile game experiences, most of the participants (80\%) play games for fun, usually $3$ -- $4$ days per week. Three of the participants consider themselves experts in playing mobile games, five have intermediate-level gaming experience, and the others are novices. Their gaming expertise is based on the number of mobile game apps they have ever played similar to our testing mobile game app and all of them have no prior knowledge of our mobile game app, i.e., nor have they seen it or played with it.

\par The mobile game app we choose for our study is a type of ARPG, where the player controls the actions of the main game character immersed in a well-defined virtual world and resists attacks from in-game monsters. The orientation and movement of the main game character are controlled by the player through a virtual joystick placed on the left side of the mobile screen, while the skill release controls are placed on the right side of the screen, usually consisting of five skills (i.e., one Normal Attack in the right-button corner surrounded by the other four skills). We have studied different mobile applications (e.g., ``address book'', ``2048'', ``Angry Birds'', ``Temple Run'') and identified that they all share the same set of basic down/move/up actions. However, ``Angry Birds'' only involves limited events of finger move (e.g., launch a bird) and finger down (e.g., make birds explode). Therefore, we choose the virtual joystick mobile game that involves lots of finger down/move/up events fully engaging players via interactions with both hands and has a proper duration ($2$ -- $5$ min on average) and different levels of difficulty. Thus, the resulting gesture interactions are diverse enough for our experimental analysis.

\subsection{Case One: Interaction Skill Comparison}
\par The objective of the first case study is to differentiate novice players from the expert ones based on their interaction data with the mobile multi-touch screen when playing our testing mobile game app. For the first case study, we recruited two novices (one female student, age: $25$ and one male student, age: $23$) and two expert gamers (male students, age: $20$ and $25$, respectively) to compare their interaction skills by using our approach. A single gaming session was conducted with each participant for $5$ minutes. They were firstly given a brief overview of the basic operating rules of the testing mobile game app. Then, each participant played for three consecutive rounds, and we only collected the operation logs of the last two attempts. The first attempt served as a training session to help the participants familiarize with the testing mobile game app.

\begin{figure*}{h}
\centering
\includegraphics[width=\linewidth]{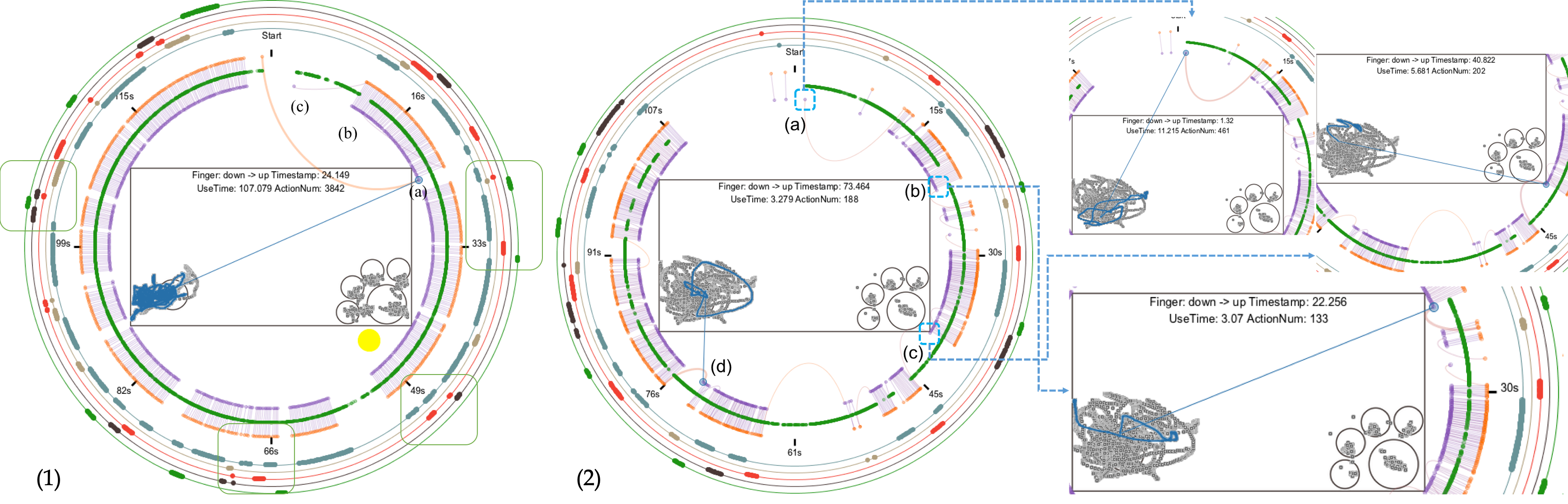}
\vspace{-6mm}
\caption{(1) Visualization result of a novice player: only three long-lasting moving trajectories were observed (a), (b), and (c). The skill combinations on the skill semantic circles indicated by the green rectangles show that the player released different skills without any strategies and did not consider the joystick interaction; (2) Visualization result of an expert player: constant moving operations were witnessed on the left side of the mobile screen ((a) -- (d)). Taking an in-depth analysis of each long-lasting trajectory of movement along with the skill release, the game expert found that after each movement, there is a combination of skills released. From the shape of gesture trajectories, they conclude that the purpose of the corresponding movement is to pull the monsters together by attracting their attention. With the ability to pull extremely well and hit the monsters with different skills, the player certainly takes down the monsters quickly and efficiently.}
  \label{fig:case1_weakgood}
    \vspace{-3mm}
\end{figure*}

\par As shown in \autoref{fig:case1_weakgood}(1), we applied our visualization approach to the interaction data of a novice player and identified that the player has only three finger movement trajectories, indicated by (a), (b), and (c), respectively. Among the three movement trajectories, (b) and (c) occur within the first $10$ seconds, followed by a long-lasting movement of over $107$ seconds that continues until the end of the game session. We then observed how the player interacted with the skill buttons by the five corresponding skill concentric semantic circles. Most of the five skill buttons are triggered simultaneously in clockwise order with the Normal Attack (the biggest button) triggered first followed by the other four skills, as indicated by the green rectangles. Since the four skills have several seconds for cooldown, they cannot be triggered immediately if they have been triggered and the Normal Attack is frequently used. We can conclude that the operation of the novice player involves a continuous movement on the left side of the screen to control the orientation and movement of the game character in the mobile game app and consecutive release of the five skills in a clockwise direction. The left (orientation and movement) and the right operation (skills) are not combined well. In other words, the player did not have a good strategy to combine the orientation control and the movement of the game character effectively with the other five skills to defend the game character against attacks.

\begin{figure}[h]
\includegraphics[width=\linewidth]{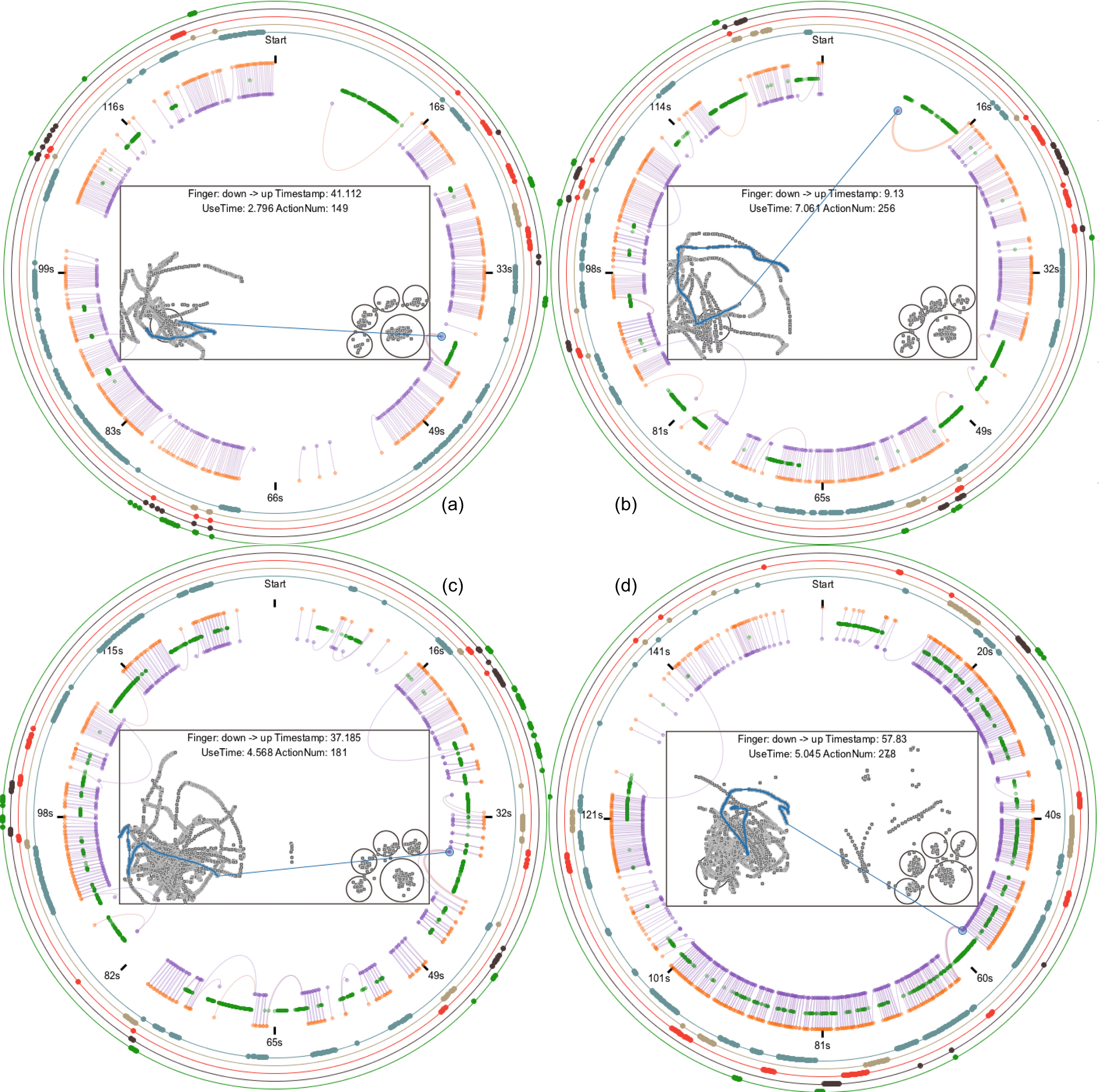}
\vspace{-6mm}
\caption{(a -- b) Continuous touchings on skill concentric circles and nearly all skill triggering points happen simultaneously. There is no correlation between movement control by joystick and skills release in the first two attempts. In the last attempt, the combination of movement control and skill release becomes obvious.}
  \label{fig:task2_com}
    \vspace{-3mm}
\end{figure}

\begin{figure*}[h]
\includegraphics[width=\linewidth]{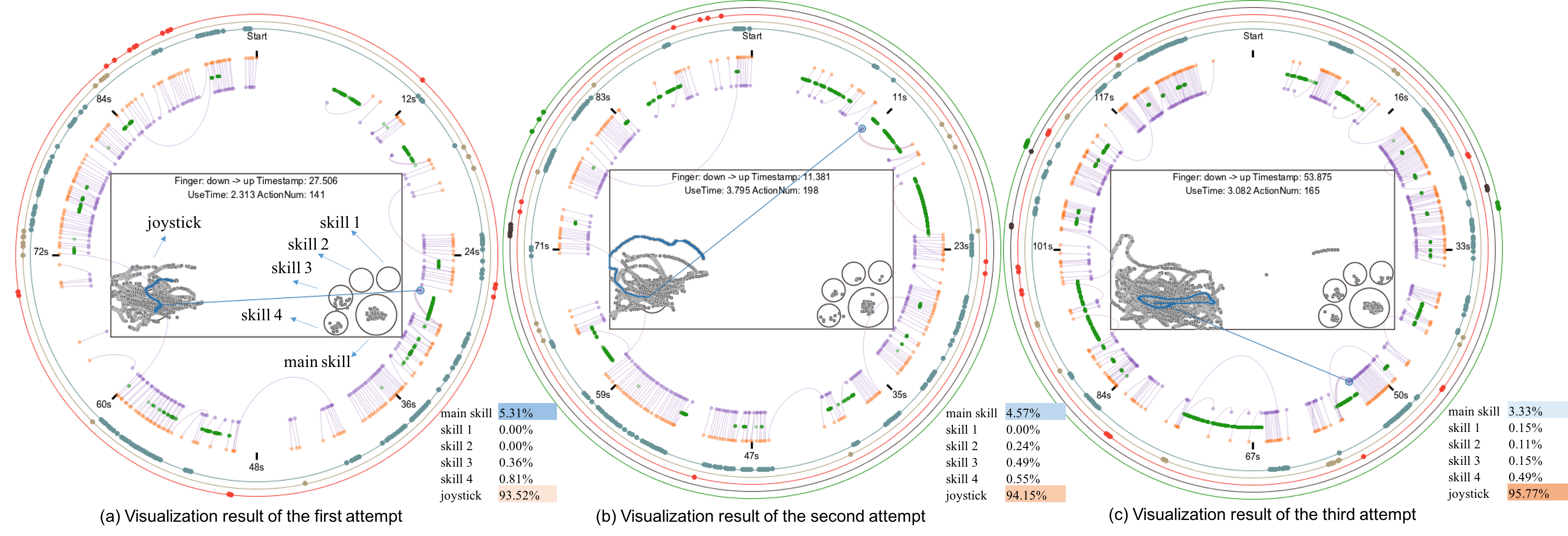}
\vspace{-6mm}
\caption{The visualization results for a series of attempts to finish a gaming task in a joystick-based mobile game application: (a) and (b) are results of failed attempts, while (c) is a successful attempt. The patterns in the series of results indicate that the participant was getting more familiar with the procedure of the mobile application, i.e., the participant gradually paid attention to the combination of the joystick with the application of different skills. When there was a long-lasting movement, some skills, apart from Normal Attack, were clicked immediately. Meanwhile, the ratio of Normal Attack significantly drops and more operations were distributed in the joystick and other four skills (joystick controls the orientation and movement of the avatar in the mobile game; Normal Attack: ordinary hit on monsters; skill 1 -- 4: different levels of hit on monsters).}
  \label{fig:teaser}
    \vspace{-3mm}
\end{figure*}

\par For comparison, we visualize another expert player's interaction data, as shown in \autoref{fig:case1_weakgood}(2). Following the same approach, we first focused on the long-last movement trajectories, which are defined by relatively long and high curves that connect the corresponding ``starting'' dots (finger down) and ``ending'' points (finger up). We identified that at the timestamp of each long-lasting movement, the skill set on the right side of the screen is triggered in a regular pattern, i.e., the Normal Attack is typically accompanied by the other four skills that require a certain cooldown time. Thus, the efficient combination of the long-lasting movement to control the orientation of the game character and the release of the skill set results in the player winning the game.

\par After comparison, the game experts maintained that the four skills, except for the Normal Attack, are lower in operations because of that ``\textit{a good operation focuses more on joystick control and movement.}'' (E.1) By contrast, a weak operation depends significantly on the Normal Attack compared with the other skills, and involves fewer finger movement on the left side of the mobile screen. From the visualizations, the game experts commented that ``\textit{the weak operation continuously utilizes the Normal Attack during the finger moving and touching operations}'' (see the most inner semantic circle in \autoref{fig:case1_weakgood}(1)) ``\textit{while the good operation applies the Normal Attack regularly and intermittently}''.

\subsection{Case Two: Individual Interaction Skill Improvement}
\par In this case, we demonstrate how the game experts leverage our visualization approach to track the potential improvement of an individual player's skills when he/she takes a series of attempts of playing a mobile game. The game experts invited another two novices players (a male student with the age of $28$ and a female student with the age of $24$) and asked them to carry out a series of attempts consecutively. Particularly, the participants were first introduced to get familiar with the basic operations and then played the mobile game several times. After finishing the attempts, their interaction data were recorded each time and the game experts conducted an interview on each of them, separately taking a note of each interviewee's response.

\par \autoref{fig:task2_com} shows a case where a participant (male, age: 28) played the game four times, of which the first two attempts failed to accomplish the game task. To be specific, in the first two attempts, although the participant applied all the skills, they were conducted successively. We observed that continuous touches occur on the second, third, and fourth skill buttons, which should be avoided since that except for the Normal Attack, all the other four skills have a cooldown time that disables the corresponding skill for a certain duration. Another phenomenon that can be witnessed is that nearly all the skill triggering happens simultaneously and there is no correlation between the movement control by the joystick and the skill operations. On the other hand, the last two results correspond to successful attempts. The combination of movement control and skill release becomes more obvious. The triggering time intervals between different skills are elongated, as shown in \autoref{fig:task2_com} (d). Furthermore, the movement trajectories are denser and more concentrated in \autoref{fig:task2_com} (d) than that in the previous attempts. ``\textit{In the beginning, I was quite nervous and did not know how to handle so many monsters so I just click on every skill buttons,}'' said the participant, ``\textit{I soon realized that without any strategy I would never win the game}'', so he began to coordinate his both hands.

\par Another participant made three attempts. In the first two attempts, the participant also failed to finish the gaming task. Particularly, in the first attempt, the participant only used Normal Attack and the other two skills. Although the movement was quite intensive, there is no combination with the application of skill release. In the second attempt, the participant began to use more skills, but in a random way. Taking a close look at the triggering timestamp of the skills and movement control, the game experts cannot identify any correlation. In the last attempt, the participant commented that ``\textit{I feel like knowing how to play}'' when he began to pay attention to the combination of the joystick with the application of different skills. For example, when there was a relatively long-lasting movement, some skills were released immediately. As shown in \autoref{fig:teaser}, the ratio of Normal Attack significantly drops, and more operations are given to the joystick and are distributed in the four skills evenly.

\begin{table*}[h]
\centering 
\caption{Summary table of touching and moving scale measurement of joystick with the confidence coefficient of 95\% and 99\%.} 
\label{table2} 
\resizebox{\textwidth}{!}{%
\begin{tabular}{|l|l|l|l|l|l|l|l|l|l|l|}
\hline
\multicolumn{11}{|c|}{Confidence Coefficient: 95\%} \\ \hline
joystick & original radius & sampling number & original center & original number & new center & new radius & new number & distance to the left & distance to the bottom & diameter \\ \hline
touch & 421 & 8908 & (1462.172, 216.930) & 1228 & (1467.603, 214.247) & 146.331 & 1166 & 15.86 mm & 13.85 mm & 15.61 mm \\ \hline
move & 444 & 87223 & (1468.190, 237.092) & 86174 & (1475.237, 234.957) & 232.275 & 81865 & 15.37 mm & 15.18 mm & 30.02 mm \\ \hline
\multicolumn{11}{|c|}{Confidence Coefficient: 99\%} \\ \hline
joystick & original radius & sampling number & original center & original number & new center & new radius & new number & distance to the left & distance to the bottom & diameter \\ \hline
touch & 421 & 8908 & (1462.172, 216.930) & 1228 & (1462.172, 216.930) & 196.650 & 1215 & 16.21 mm & 14.02 mm & 25.42 mm \\ \hline
move & 444 & 87223 & (1468.190, 237.092) & 86174 & (1470.297, 236.474) & 321.171 & 85312 & 15.68 mm & 15.28 mm & 41.51 mm \\ \hline
\end{tabular}%
}
\end{table*}

\subsection{Case Three: UI Design Verification}
\par This case focuses on identifying the common behaviors of players' interaction with the mobile game app and further verifying the current UI design of the mobile app interface in real usage scenarios. To approach this, the game experts covered the main operation regions of the mobile game app interface to ensure that the participants operate based on their own habits for an objective evaluation. As shown in \autoref{fig:procedure1}(b), the size of the covered region is approximately $3$ cm * $3.5$ cm. The game experts invited all the participants and asked them to use only two thumbs to operation on the game avatar in the mobile game app for five rounds, no matter success or failure for accomplishing the gaming task. During the gaming process, all the participants were asked to keep their thumbs in a natural curve and use their finger bellies to touch the screen while maintaining the stability of the mobile devices. The mobile device used has a width of $11.07$ cm and a height of $6.23$ cm. The resolution of the device was $1920$ * $1080$ pixels and the OS is Android. In total, we recorded $45$ independent interaction data, in which about $400$ gesture trajectories were considered valid. After the experiments, the game experts conducted an interview with the participants to collect their subjective feedback when playing the mobile game app.

\par
\begin{figure}[h]
\centering
\includegraphics[width=\linewidth]{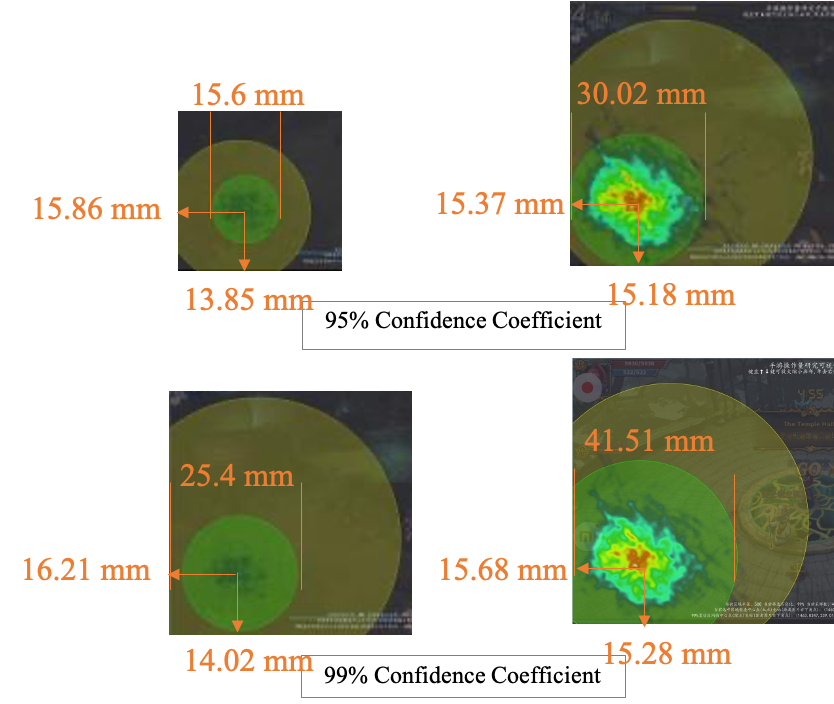}
\vspace{-6mm}
\caption{Distances to the left and bottom and diameter of joystick when the participants touched and moved freely with 95\% and 99\% confidence coefficient.}
  \label{fig:joystick_heat}
  \vspace{-3mm}
\end{figure}

 \textbf{Virtual Joystick UI Verification.} The game experts first focused on the triggering area and moving region of the joystick on the left side of the mobile screen. We applied interactive clustering on all the touching sampling dots generated by the participants' interactions. To obtain the minimum coverage of the triggering area of the virtual joystick, we covered the interaction area with a yellow circle with the radius of $R_0$. Through the interactive clustering, a new radius $R_{min}$ is determined. To obtain the upper boundary of the triggering area, we widened the initial coverage and adjusted the confidence coefficient to 99\%. Another new radius $R_{max}$ was observed. Meanwhile, the distance to the left boundary and the bottom boundary is recorded to locate the specific position of the new center. The experimental results indicate that the lower and upper boundary of the triggering area is $15.61$ mm and $25.42$ mm based on our method and the distances to the left and the bottom boundary of the screen were $15.86$ mm and $13.85$ mm with the confidence coefficient of $95\%$. The same procedure can be applied to determine the moving range of the joystick. We summarize our findings in Table~\ref{table2} and \autoref{fig:joystick_heat}.

\par
\begin{figure*}[h]
\includegraphics[width=\linewidth]{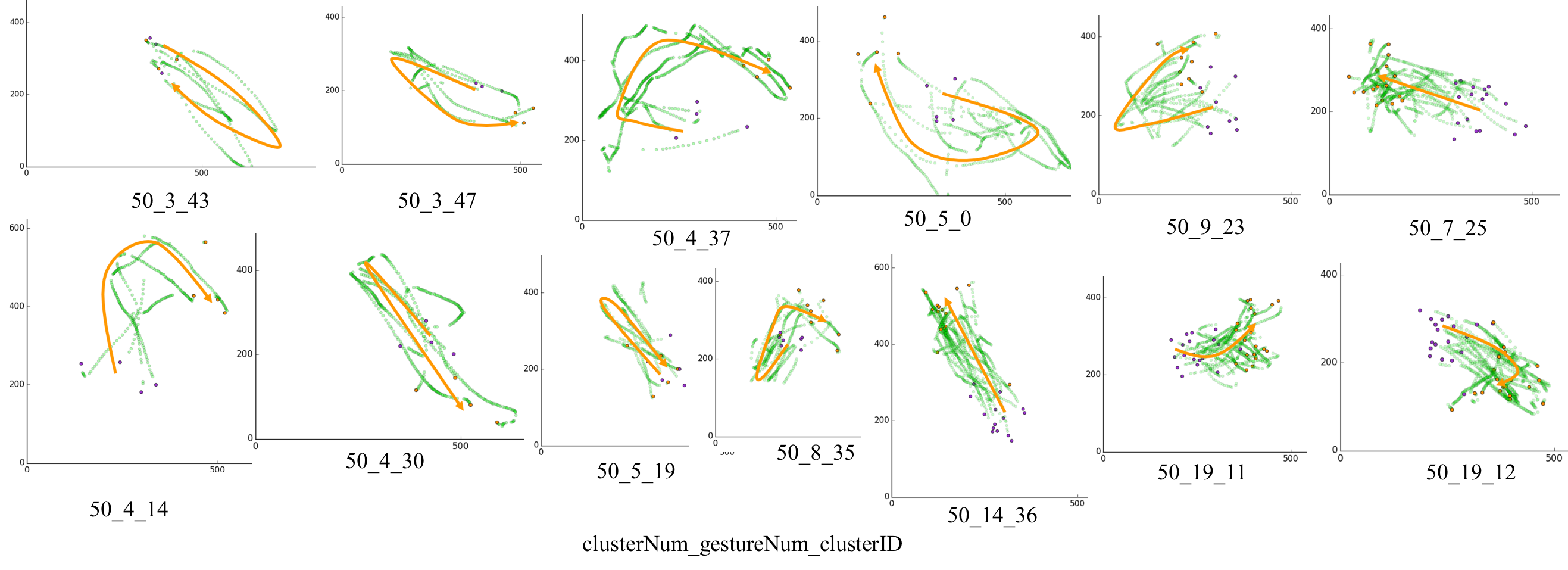}
  \vspace{-6mm}
\caption{Summary of gesture trajectories that demand fingers to move a relatively long distance.}
  \label{fig:cluster}
    \vspace{-3mm}
\end{figure*}

 To further identify the major trajectories using the virtual joystick, we clustered the gestures that lie within the boundary of the moving area of the virtual joystick based on the proposed clustering method. Eexperimental results show the patterns that occupy the majority of the trajectories. To be specific, apart from those gestures with short distances, the gestures with relatively longer distances are clustered into $13$ clusters, representing the main gestures that demand high workload, i.e., moving fingers with a relatively long distance. As shown in \autoref{fig:cluster}, the new boundary of moving area of the virtual joystick has covered the most frequent gestures that take a relatively long distance for fingers to move, which indicates that the above experimental scale suggestions can support a free movement according to the participants' operation habits.

\par
\begin{figure}[h]
\includegraphics[width=\linewidth]{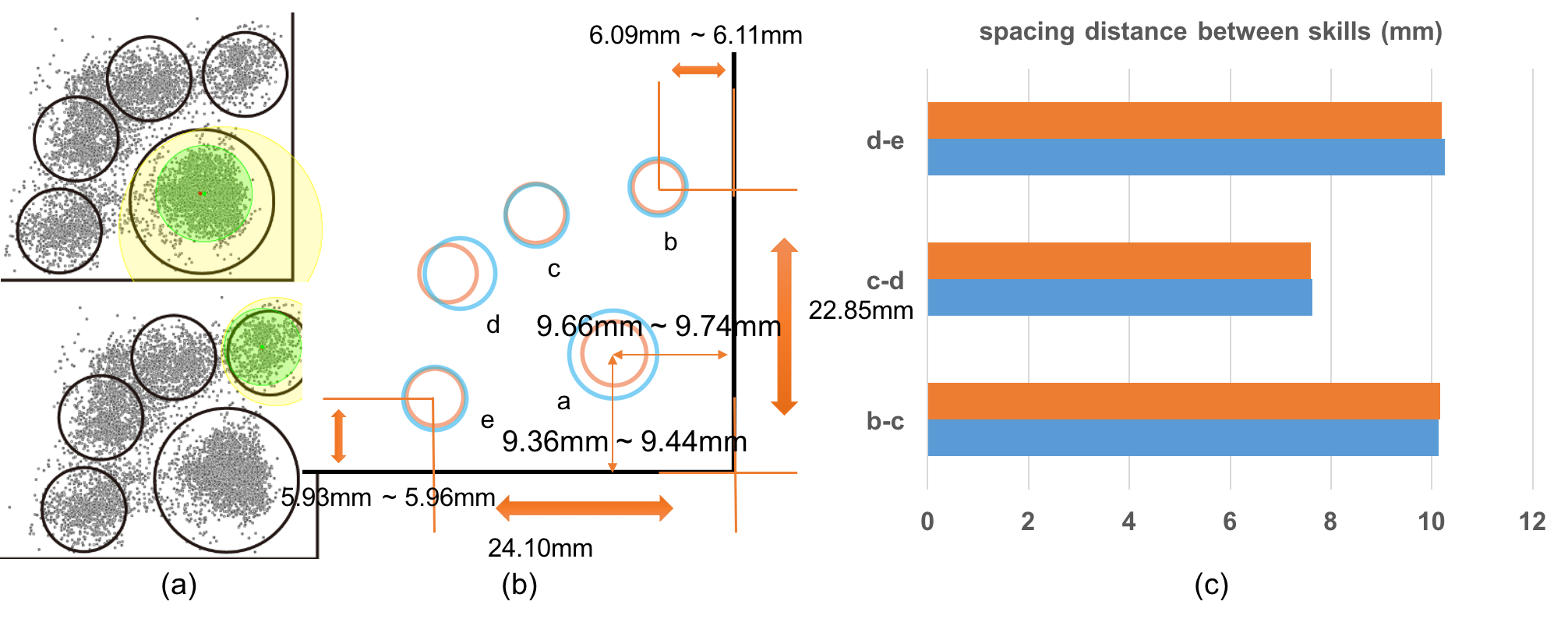}
  \vspace{-6mm}
\caption{(a) Screen area clustering of skill set; (b) the distances from the skill set to the boundaries of the mobile screen; (c) the spacing distance between skill buttons through our experiment.}
  \label{fig:skillset}
    \vspace{-3mm}
\end{figure}

 \textbf{Skill Set UI Verification.} Following a similar procedure, the game experts determined the scale and position of the skill responding area by adopting interactive clustering that covers all the touching sampling dots in the skill set area. Table~\ref{table3} gives a summary result of the experimental results for the skill set. Particularly, with the confidence coefficient range of $90\%$ and $95\%$, the spacing among skill buttons is in the range of $7.6$ mm -- $10.25$ mm. Due to the fact that players may easily confuse with the middle two skill buttons, the spacing should be relatively larger between c and d. Similarly, the distances between the centers of the skill buttons and the right/bottom boundary of the mobile screen are in the range of $6.09$ mm -- $24.10$ mm and $5.93$ mm -- $22.85$ mm, while the distances between the center of the Normal Attack and the right/bottom boundary are in the range of $9.66$ mm -- $9.74$ mm and $9.36$ mm -- $9.44$ mm, as indicated in \autoref{fig:skillset}.

\begin{table*}[h]
\centering 
\caption{Summary table of touching and moving scale measurement of joystick with the confidence coefficient of 95\% and 99\%.} 
\label{table3} 
\resizebox{\textwidth}{!}{%
\begin{tabular}{lllllllllll}
\hline
\multicolumn{11}{|c|}{\textbf{Confidence Coefficient: 90\%}} \\ \hline
\multicolumn{1}{|l|}{\textbf{skill set}} & \multicolumn{1}{l|}{\textbf{original radius}} & \multicolumn{1}{l|}{\textbf{sampling number}} & \multicolumn{1}{l|}{\textbf{original center}} & \multicolumn{1}{l|}{\textbf{original number}} & \multicolumn{1}{l|}{\textbf{new center}} & \multicolumn{1}{l|}{\textbf{new radius}} & \multicolumn{1}{l|}{\textbf{new number}} & \multicolumn{1}{l|}{\textbf{distance to the left}} & \multicolumn{1}{l|}{\textbf{distance to the bottom}} & \multicolumn{1}{l|}{\textbf{diameter}} \\ \hline
\multicolumn{1}{|l|}{\textbf{a}} & \multicolumn{1}{l|}{169} & \multicolumn{1}{l|}{8908} & \multicolumn{1}{l|}{(156.415, 146.934)} & \multicolumn{1}{l|}{3183} & \multicolumn{1}{l|}{(149.757, 146.309)} & \multicolumn{1}{l|}{79.890} & \multicolumn{1}{l|}{2864} & \multicolumn{1}{l|}{9.66 mm} & \multicolumn{1}{l|}{9.44 mm} & \multicolumn{1}{l|}{5.15 mm} \\ \hline
\multicolumn{1}{|l|}{\textbf{b}} & \multicolumn{1}{l|}{90} & \multicolumn{1}{l|}{8908} & \multicolumn{1}{l|}{(94.726, 353.550)} & \multicolumn{1}{l|}{747} & \multicolumn{1}{l|}{(94.396, 354.230)} & \multicolumn{1}{l|}{63.702} & \multicolumn{1}{l|}{672} & \multicolumn{1}{l|}{6.09 mm} & \multicolumn{1}{l|}{22.85 mm} & \multicolumn{1}{l|}{4.11 mm} \\ \hline
\multicolumn{1}{|l|}{\textbf{c}} & \multicolumn{1}{l|}{80} & \multicolumn{1}{l|}{8908} & \multicolumn{1}{l|}{(246.772, 321.063)} & \multicolumn{1}{l|}{1263} & \multicolumn{1}{l|}{(247.711, 321.218)} & \multicolumn{1}{l|}{74.087} & \multicolumn{1}{l|}{1136} & \multicolumn{1}{l|}{15.98 mm} & \multicolumn{1}{l|}{20.72 mm} & \multicolumn{1}{l|}{4.78 mm} \\ \hline
\multicolumn{1}{|l|}{\textbf{d}} & \multicolumn{1}{l|}{80} & \multicolumn{1}{l|}{8908} & \multicolumn{1}{l|}{(341.600, 243.923)} & \multicolumn{1}{l|}{1341} & \multicolumn{1}{l|}{(339.483, 247.096)} & \multicolumn{1}{l|}{71.902} & \multicolumn{1}{l|}{1206} & \multicolumn{1}{l|}{21.90 mm} & \multicolumn{1}{l|}{15.94 mm} & \multicolumn{1}{l|}{4.64 mm} \\ \hline
\multicolumn{1}{|l|}{\textbf{e}} & \multicolumn{1}{l|}{80} & \multicolumn{1}{l|}{8908} & \multicolumn{1}{l|}{(375.057, 92.975)} & \multicolumn{1}{l|}{1075} & \multicolumn{1}{l|}{(372.814, 91.963)} & \multicolumn{1}{l|}{71.109} & \multicolumn{1}{l|}{967} & \multicolumn{1}{l|}{24.05 mm} & \multicolumn{1}{l|}{5.93 mm} & \multicolumn{1}{l|}{4.59 mm} \\ \hline
 &  &  &  &  &  &  &  &  &  &  \\ \hline
\multicolumn{11}{|c|}{\textbf{Confidence Coefficient: 95\%}} \\ \hline
\multicolumn{1}{|l|}{\textbf{skill set}} & \multicolumn{1}{l|}{\textbf{original radius}} & \multicolumn{1}{l|}{\textbf{sampling number}} & \multicolumn{1}{l|}{\textbf{original center}} & \multicolumn{1}{l|}{\textbf{original number}} & \multicolumn{1}{l|}{\textbf{new center}} & \multicolumn{1}{l|}{\textbf{new radius}} & \multicolumn{1}{l|}{\textbf{new number}} & \multicolumn{1}{l|}{\textbf{distance to the left}} & \multicolumn{1}{l|}{\textbf{distance to the bottom}} & \multicolumn{1}{l|}{\textbf{diameter}} \\ \hline
\multicolumn{1}{|l|}{\textbf{a}} & \multicolumn{1}{l|}{169} & \multicolumn{1}{l|}{8908} & \multicolumn{1}{l|}{(156.415, 146.934)} & \multicolumn{1}{l|}{3183} & \multicolumn{1}{l|}{(150.972, 145.146)} & \multicolumn{1}{l|}{109.730} & \multicolumn{1}{l|}{3023} & \multicolumn{1}{l|}{9.74 mm} & \multicolumn{1}{l|}{9.36 mm} & \multicolumn{1}{l|}{7.08 mm} \\ \hline
\multicolumn{1}{|l|}{\textbf{b}} & \multicolumn{1}{l|}{90} & \multicolumn{1}{l|}{8908} & \multicolumn{1}{l|}{(94.726, 353.550)} & \multicolumn{1}{l|}{747} & \multicolumn{1}{l|}{(94.782, 354.228)} & \multicolumn{1}{l|}{71.214} & \multicolumn{1}{l|}{709} & \multicolumn{1}{l|}{6.11 mm} & \multicolumn{1}{l|}{22.85 mm} & \multicolumn{1}{l|}{4.59 mm} \\ \hline
\multicolumn{1}{|l|}{\textbf{c}} & \multicolumn{1}{l|}{80} & \multicolumn{1}{l|}{8908} & \multicolumn{1}{l|}{(246.772, 321.063)} & \multicolumn{1}{l|}{1263} & \multicolumn{1}{l|}{(248.461, 320.519)} & \multicolumn{1}{l|}{77.606} & \multicolumn{1}{l|}{1199} & \multicolumn{1}{l|}{16.03 mm} & \multicolumn{1}{l|}{20.68 mm} & \multicolumn{1}{l|}{5.01 mm} \\ \hline
\multicolumn{1}{|l|}{\textbf{d}} & \multicolumn{1}{l|}{80} & \multicolumn{1}{l|}{8908} & \multicolumn{1}{l|}{(341.600, 243.923)} & \multicolumn{1}{l|}{1341} & \multicolumn{1}{l|}{(339.952, 246.550)} & \multicolumn{1}{l|}{88.189} & \multicolumn{1}{l|}{1273} & \multicolumn{1}{l|}{21.93 mm} & \multicolumn{1}{l|}{15.91 mm} & \multicolumn{1}{l|}{5.69 mm} \\ \hline
\multicolumn{1}{|l|}{\textbf{e}} & \multicolumn{1}{l|}{80} & \multicolumn{1}{l|}{8908} & \multicolumn{1}{l|}{(375.057, 92.975)} & \multicolumn{1}{l|}{1075} & \multicolumn{1}{l|}{(373.580,92.420)} & \multicolumn{1}{l|}{78.870} & \multicolumn{1}{l|}{1021} & \multicolumn{1}{l|}{24.10 mm} & \multicolumn{1}{l|}{5.96 mm} & \multicolumn{1}{l|}{5.09 mm} \\ \hline
\end{tabular}%
}
\end{table*}

\section{Discussion and Reflections}
\subsection{System Performance and Generality}
\par We first asked the game experts to evaluate the insights identified by our visualization approach. E.1 -- 2 reported that ``\textit{the visualization displays the interaction patterns of players intuitively}''. Conventionally, the game team needed to watch video replays, and manually marked segments of interactions: ``\textit{the time we spent on the entire process was about $30$ minutes since sometimes we had to replay a certain video segment many times,}'' said E.1. Our method can visualize the spatiotemporal attributes of the gesture interactions, making it easy for the experts to interact with players' behaviors with the mobile game app instead of going through the entire video replay, which largely shortens the analysis time (now around $5$ minutes for each session). The UI design experts were very satisfied with our approach's ability to spot potential UI design issues for the joystick and skill areas, which serves as a complementary to traditional methods such as survey other game competitors, ``\textit{it helps me obtain detailed UI suggestions by interacting the interface,}'' said E.4, ``\textit{I am now more confident to draw my UI design conclusions since my subjective feelings can be confirmed,}'' said E.5.
\par We also discussed with the five game experts which component(s) of our approach can be directly applied to other kinds of games or apps and which one(s) need customization. They all appreciated that our approach has been already applicable to any kind of ARPG apps since their functions are quite unified, mainly around joysticks and skills. The only place that needs customization is the middle part of the interface, i.e., the mobile game app interface that provides gaming contexts.

\subsection{Design Implications for Mobile Game App UI}
\par Our visualizations have provided some design implications for the UI design of all kinds of ARPG apps. Players spend most of the time operating with the joystick, which is also the most energy-consuming; therefore, it has the highest requirements for flexibility and fineness. ``\textit{The purpose of the virtual joystick is to control the distance and direction of movement and the orientation of the game avatar,}'' said E.4. Participants also reported their requirements for the virtual joysticks include: 1) flexible and fast control of the position of the game avatar; 2) precise control of the distance and direction of movement; 3) lasting operation does not consume too much energy. The results from our visualizations also indicate that the finger movement areas should be sufficient enough to ensure that most areas of the mobile screen are responsive, ``\textit{it is best to move anywhere you want,}'' said one participant (P1, male, age: 25).

\par In terms of the skill interaction design, UI design experts reported that their design principles are to support clicking on the graphical representations of the buttons to release skills and to display cooldowns when the skills are not available. From the perspective of participants, they wanted the design of the skill buttons to meet the following requirements: 1) to release skills quickly and easily; 2) to accurately release skills with less attention; 3) to quickly know when skills will be available. 
Through the quantitative experimental analysis of participants' interaction data, UI design experts found that participants are easy to locate high-frequency buttons through the edge of the screen, which should be located along the ``fan-shaped'' curve. Other lower-frequency operation buttons such as switching roles can be placed in the area near the $41.3$ mm arc, facilitating easy clicking and locating. In addition, the participants reported that the middle two skill buttons are more easily noticed during operation and are suitable for major/important skills, while skills near the border on both sides are relatively more suitable for auxiliary skills such as dodging.

\par Our game experts also pointed out several directions for improving our visualization approach. E.1 and E.2 hoped that we could take in the in-game video and metrics from the mobile game app as a whole, enabling a comprehensive analysis. It is for ensuring the high ``\textit{consistency of analysis conclusion}.'' Meanwhile, the game experts commented that ``{the method has the potential to be developed into a training tool to support a retrospective analysis of players' performance.}'' The UI design experts (E.4 -- 5) also hoped that we could include more in-game contents. For example, a proper number of monsters and stable in-game fighting duration together with appropriate UI design can not only prevent the players from ``\textit{being in a state of high operating frequency all the time}'', but also ``\textit{bring the players a sense of satisfaction}''.

\section{Conclusion}
\par In this work, we introduce a visualization approach to explore the interaction data of players on multi-touch mobile devices. The interaction data is transformed to gesture-based data and new visualization techniques are integrated to ease the exploration of interaction data. Three case studies and feedback from the game experts confirm the efficacy of our approach. In the future, we plan to include longitudinal studies to validate our approach and include other kinds of mobile game applications to better spot the limitations. Furthermore, embedding the game video in our visualization approach is a promising way to better learn how players behave.

\begin{acks}
We are grateful for the valuable feedback and comments provided by the anonymous reviewers. This work is partially supported by the research start-up fund of ShanghaiTech University and HKUST-WeBank Joint Laboratory Project Grant No.: WEB19EG01-d.
\end{acks}
\balance
\bibliographystyle{ACM-Reference-Format}
\bibliography{sample-base}


\end{document}